\documentstyle[12pt]{article}
\advance\textheight by \topskip
\begin{document}
\begin{flushright}
SNUTP-97-031\\
February 18, 1997\\
\end{flushright}
\vspace{1 cm}
\begin{center}
\baselineskip=16pt

{\Large\bf  Quantum Potential and Cosmological Singularities}

\vskip 2cm
{\bf Sang Pyo Kim}\footnote{E-mail:
sangkim@knusun1.kunsan.ac.kr}\\
\vskip 0.8cm

Department of Physics\\
Kunsan National University \\
Kunsan 573-701, Korea\\

\vskip 1 cm

\end{center}

\vskip 1 cm

\centerline{\bf ABSTRACT}
\begin{quotation}
We apply the de Broglie-Bohm interpretation to the Wheeler-DeWitt equation
for the quantum FRW cosmological model with a minimal massless scalar field.
We find that the quantum FRW cosmological model has quantum potential
dominated solutions that avoid the initial and the final cosmological
singularities. It is suggested that the quantum potential and
the back-reaction of geometry and matter fields may change the property of
the cosmological singularities of the Universe.
\end{quotation}

\newpage

\section{Introduction}

One of the most prominent classical-mechanical problems that
quantum mechanics can solve is the classical singularity.
For example, an electron of an atom suffers from the instability that due to
the infinitely negative Coulomb potential
and the electron in a zero-momentum state should fall into the
infinite potential\footnote{More rigorously, one should take
into account the energy loss due to the radiation of the electron
which renders it to fall finally into the nucleus}.
In quantum mechanics, however, the $s$-wavefunction (zero orbital
angular momentum state)
of the electron has zero probability at the singularity of potential.

Similarly, according to the classical singularity theorem by Hawking
and Penrose, the Universe also suffers from the initial and the
final singularities \cite{hawking}.
One of the many prospects of canonical quantum gravity
which is described by the Wheeler-DeWitt equation
from the Hamiltonian formulation of
the general relativity \cite{dewitt} whose wavefunctions
carry all the information of the gravity and matter system,
has been to resolve the classical singularities, as quantum mechanics does.
In spite of diverse approaches and attempts \cite{blyth,gotay,lemos,peleg},
the problem of singularities for gravity coupled to a general matter field
has not been completely solved yet.

In this Letter we approach the problem of the initial and the final
singularities of the Universe by applying the de Broglie-Bohm interpretation
to the wavefunctions of the Wheeler-DeWitt equation for the quantum FRW
cosmological model with a minimal massless scalar field.
The de Broglie-Bohm interpretation is an alternative approach to
quantum mechanics, in which the Schr\"{o}dinger equation is
replaced by two equations, the Hamilton-Jacobi equation
with the quantum potential and the continuity equation for
probability \cite{holland}. The quantum trajectory described by
these two equations are physically the same as the corresponding
wavefunction of Schr\"{o}dinger equation.
By introducing a suitable
cosmological time-parameter and solving the imaginary part of
the Wheeler-DeWitt equation (the equation for the continuity
of probability in quantum mechanics), we are able to find the semiclassical
Einstein equation both with the quantum potential as a part of quantum
fluctuation of geometry and with an operator ordering parameter.
We find the solutions for the semiclassical Einstein. Depending on the
relative magnitude of the operator ordering parameter and the energy
density of the scalar field, there can be or cannot be particular
solutions that avoid the cosmological singularities, either directly via
the exact wavefunctions or via the de Broglie-Bohm interpretation.

The exact solvability of the de Broglie-Bohm trajectory
for the quantum FRW model is rooted on the direct separability
of the Wheeler-DeWitt equation into the gravitational field equation
by quantum states of the massless scalar field.
For a general scalar-field model, one may need the semiclassical
quantum gravity. Several different ways have been introduced
to derive the semiclassical quantum gravity \cite{kiefer}.
Firstly, one may use the Born-Oppenheimer idea of separating
the different mass scales to get the semiclassical
Einstein equation \cite{brout}; secondly, one may apply
the de Broglie-Bohm idea to the Wheeler-DeWitt equation \cite{vink};
thirdly, one may use both the Born-Oppenheimer and the de Broglie-Bohm
ideas \cite{kim1,kim2}.

\section{Classical and Quantum FRW Model}

We shall consider the FRW cosmological model minimally coupled to
a massless scalar field. The action for the gravity
coupled to the minimal massless scalar field is given by
\begin{equation}
I = - \frac{m_P}{16 \pi} \int d^4 x \sqrt{-g} R
+ \frac{1}{2} \int d^4 \sqrt{-g} g^{\mu \nu}
\partial_{\mu}\phi \partial_{\nu}
\phi.
\end{equation}
In the FRW geometry with the metric
\begin{equation}
ds^2 = - N^2(t) dt^2 + a^2 d \Omega_3^2,
\end{equation}
the action takes the form
\begin{equation}
I = \int dt \Biggl[ - \frac{3m_P}{8 \pi}
a^3 \Bigl( \frac{\dot{a}^2}{Na^2}
 - N \frac{k}{a^2} \Bigr)
+  \frac{a^3 \dot{\phi}^2}{2 N}  \Biggr],
\label{act}
\end{equation}
where $k = 1, 0, -1$ for a spatially closed, flat,
open Universe, respectively.

{}From the action (\ref{act}), one obtains the classical Einstein equation
\begin{equation}
\Bigl( \frac{\dot{a}}{a} \Bigr)^2 + \frac{k}{a^2}
=  \frac{4 \pi}{3 m_P} \dot{\phi}^2,
\label{ein eq}
\end{equation}
and the classical equation for the scalar field
\begin{equation}
\ddot{\phi} + 3 \frac{\dot{a}}{a} \dot{\phi} = 0.
\label{cl f}
\end{equation}
As a simple case, we shall consider the massless scalar field
in the spatially flat Universe. Eq. (\ref{cl f}) has the solution
\begin{equation}
\dot{\phi} = \frac{p}{a^3}
\end{equation}
where $p$ is a constant of integration (classical momentum),
and the corresponding solutions of Eq. (\ref{ein eq}) are
\begin{equation}
a(t) = a(t_0) \Bigl[ 1 \pm
\sqrt{\frac{4 \pi p^2}{m_P}} \frac{3}{a^3(t_0)}
\bigl( t - t_0 \bigr) \Bigr]^{1/3}.
\end{equation}
The positive and the negative signs correspond to
the expanding and the recollapsing Universes, respectively.
Classically the Universe recollapses inevitably to the final singularity
(big crunch) or has the initial singularity run backward in time.

The Wheeler-DeWitt equation for the quantum FRW cosmological model takes
the form
\begin{equation}
\Biggl[ -
 \frac{\hbar^2}{2 m_P} \frac{1}{a^{\nu}}
\frac{\partial}{\partial a}\Bigl(a^{\nu}
\frac{\partial}{\partial a} \Bigr)
+ \frac{9m_P}{32 \pi^2} ka^2 + \frac{3 \hbar^2}{8 \pi a^2}
\frac{\partial^2}{\partial \phi^2} \Biggr] \Psi(a, \phi) = 0.
\end{equation}
In the above equation, $\nu$, which denotes an operator ordering parameter,
may have a special meaning that one can choose a particular value in
order to make the super-Hamiltonian operator hermitian in the range
$[0, \infty)$ in which the wavefunctions are defined.
In Refs. \cite{blyth,gotay,lemos} an intensive investigation
is done for the physical meaning of the parameter
$\nu$ in defining the Hilbert space of the wavefunctions.

\section{de Broglie-Bohm Interpretation}

The massless scalar-field model is a particularly simple case whose
solutions can be found.
The scalar field decouples from gravity, and  the wavefunction
has the form
\begin{equation}
\Psi(a, \phi) = \Psi(a) \Phi (\phi, a).
\end{equation}
The scalar-field Hamiltonian has the eigenfunction
\begin{equation}
\Phi = \frac{1}{(2 \pi)^{3/2}} e^{ \pm i p \phi}.
\end{equation}
Then the Wheeler-DeWitt equation reduces to the gravitational
field equation
\begin{equation}
\Biggl[ -  \frac{\hbar^2}{2 m_P} \frac{1}{a^{\nu}}
\frac{\partial}{\partial a}\Bigl(a^{\nu} \frac{\partial}{\partial a} \Bigr)
+ \frac{9 m_P}{32 \pi^2} ka^2 -  \frac{3\hbar^2 p^2}{8 \pi a^2}
\Biggr] \Psi(a) = 0.
\label{gra}
\end{equation}

One can  either solve the Wheeler-DeWitt equation directly
or follow the de Broglie-Bohm interpretation.
We shall use first the semiclassical
gravity and then the exact wavefunctions to test
the validity of it. Following de Broglie and Bohm \cite{holland},
we find the wavefunction of the form
\begin{equation}
\Psi (a) = F(a) \exp \Bigl( \pm \frac{i}{\hbar} S (a) \Bigr)
\label{db}
\end{equation}
where the positive and the negatives signs correspond
to the expanding and the recollapsing Universes, respectively.
By equating the real part, we obtain the semiclassical Einstein
(Hamilton-Jacobi) equation
\begin{equation}
 \frac{1}{2m_P} \Bigl(\frac{\partial S}{\partial a} \Bigr)^2
+ \frac{9m_P}{32 \pi^2} k a^2 - \frac{3 \hbar^2 p^2}{8 \pi a^2}
+ V_{q} (a) = 0,
\label{real}
\end{equation}
where
\begin{equation}
V_{q} (a) =
- \frac{\hbar^2}{2m_P} \Biggl[
 \frac{\frac{\partial^2 F}{\partial a^2}}{F}+
\frac{\nu}{a} \frac{\frac{\partial F}{\partial a}}{F}
\Biggr]
\label{quant}
\end{equation}
is the quantum potential.
The imaginary part leads to the continuity equation for probability
\begin{equation}
F \frac{\partial^2 S}{\partial a^2}
+ 2 \frac{\partial F}{\partial a}
\frac{\partial S}{\partial a} + \frac{\nu}{a}
F \frac{\partial S}{\partial a} = 0.
\label{imag}
\end{equation}
A cosmological time emerges into the classical world as
parameterizing the de Broglie-Bohm trajectory
along the tangential direction of the gravitational action:
\begin{equation}
\frac{\partial}{\partial t}
= \mp \frac{4\pi}{3 m_P a} \frac{\partial S}{\partial a}
\frac{\partial}{\partial a}.
\end{equation}
It is not difficult to see that if one neglects the quantum potential
and identifies the classical momentum by $p_c = \hbar p$, then
the time parameter coincides with the cosmological time used to
define the momentum in Hamiltonian formulation of the classical gravity:
\begin{equation}
\pi_a = m_P \dot{a} = \mp \frac{4 \pi}{3a}
\frac{\partial S}{\partial a}.
\label{mom}
\end{equation}
We rewrite the semiclassical Einstein equation (\ref{real}) as
\begin{equation}
\Bigl( \frac{\dot{a}}{a} \Bigr)^2
+ \frac{k}{a^2} = \frac{16 \pi^2 \hbar^2}{9m_P a^6}
\Bigl( \frac{3}{4 \pi} p^2 - \frac{2 a^2}{\hbar^2} V_{q}(a) \Bigr).
\end{equation}
Assuming a one-to-one mapping between $t$ and $a$ and using Eq. (\ref{mom}),
we are able to invert $S(a)$ and $F(a)$ as  functions
of $a(t)$ to find
\begin{equation}
V_{q} = - \frac{\hbar^2}{2m_P} \Biggl[
 \Bigl( \frac{\bigl(a \dot{a}\bigr)^{\cdot}+ \nu \dot{a}^2}{2 a
\dot{a}^2} \Bigr)^2 - \frac{1}{2 \dot{a}}
\Bigl(\frac{\bigl(a \dot{a} \bigr)^{\cdot} + \nu \dot{a}^2}{a \dot{a}^2}
\Bigr)^{\cdot} - \frac{\nu}{2}
 \frac{\bigl(a \dot{a}\bigr)^{\cdot}+ \nu \dot{a}^2}{a^2
\dot{a}^2 } \Biggr].
\end{equation}
Finally, we obtain the semiclassical Einstein equation
\begin{equation}
\Bigl( \frac{\dot{a}}{a} \Bigr)^2
+ \frac{k}{a^2} = \frac{16 \pi^2 \hbar^2}{9m_P^2}
\frac{1}{a^6}
\Biggl[  \frac{3 m_P p^2}{4 \pi}
+  a^2  \Biggl(
 \Bigl( \frac{\bigl(a \dot{a}\bigr)^{\cdot}+ \nu \dot{a}^2}{2 a
\dot{a}^2 } \Bigr)^2 - \frac{1}{2 \dot{a}}
\Bigl(\frac{\bigl(a \dot{a} \bigr)^{\cdot} + \nu \dot{a}^2}{a \dot{a}^2}
\Bigr)^{\cdot} - \frac{\nu}{2}
 \frac{\bigl(a \dot{a}\bigr)^{\cdot}+ \nu \dot{a}^2}{a^2
\dot{a}^2 }\Bigr) \Biggr].
\label{sem eq}
\end{equation}

Eq. (\ref{sem eq}) is highly nonlinear. In the flat space $ ( k = 0)$,
however, the massless scalar-field  model
has a relatively simple solution. Trying an ansatz
\begin{equation}
a(t) = a(t_0) \Bigl[ 1 \pm \beta \bigl( t- t_0 \bigr) \Bigr]^{1/3},
\end{equation}
simplifies the quantum potential as
\begin{equation}
V_{q} (a) = \frac{\hbar^2}{2m_P} \frac{(1-\nu)^2}{4 a^2},
\end{equation}
and leads to the semiclassical Einstein equation
\begin{equation}
\Bigl( \frac{\dot{a}}{a} \Bigr)^2
 = \frac{16 \pi^2 \hbar^2}{9m_P^2}
\frac{1}{a^6}
\Biggl[  \frac{3 m_P p^2}{4 \pi} - \frac{(1-\nu)^2}{4}\Biggr].
\end{equation}
Letting
\begin{equation}
D_{\nu} = \frac{3 m_P p^2}{4 \pi} - \frac{(1-\nu)^2}{4},
\end{equation}
we see that depending on the signs of $D_{\nu}$,
there either can be or cannot be classically allowed solution for $a(t)$.
For a positive $D_{\nu}$, we find the solutions
\begin{equation}
a(t) =  a(t_0) \Bigl[ 1  \pm \frac{4 \pi \hbar \sqrt{D_{\nu}}}{3m_P a^3(t_0)}
(t - t_0 ) \Bigr]^{1/3}.
\label{exact}
\end{equation}
For a negative $D_{\nu}$, there are no classically allowed solutions.
It is found that the role of the quantum potential is
to change the property of singularities.
Since the wavefunctions do not oscillate but behave  as a power-law,
we may interpret the corresponding wavefunctions as
quantum mechanical wormholes \cite{page}.

We now turn to the exact wavefunctions of the Wheeler-DeWitt equation.
We treat separately the cases $ k = 0, 1, -1$.

\subsection{k = 0}

For the case $k = 0$, we find the exact wavefunctions of the form
\begin{equation}
\Psi (a) = C \exp \bigl( \gamma \ln (a) \bigr),
\end{equation}
where $\gamma$ satisfies
\begin{equation}
\gamma^2 + (p-1) \gamma +  \frac{3 m_P p^2}{4 \pi}
= 0.
\end{equation}
Again depending on the signs of $D_{\nu}$,
the wavefunctions either oscillate or do not as the Universe recollapses.
In the case $D_{\nu} > 0$, the wavefunctions are
\begin{equation}
\Psi^{\pm}_{I} (a) = C_{\pm} a^{(1-p)/2} e^{\pm i \sqrt{D_{\nu}} \ln (a)},
\end{equation}
whereas in the case $D_{\nu} < 0$ they are
\begin{equation}
\Psi^{\pm}_{II} (a) = C_{\pm} a^{(1-p)/2 \pm \sqrt{- D_{\nu}}}.
\end{equation}
One sees that for the exact wavefunctions with the action
\begin{equation}
S = \pm \hbar \sqrt{D_{\nu}} \ln (a)
\end{equation}
Eq. (\ref{imag}) and Eq. (\ref{real}) are satisfied.
It should be noted that
the wavefunctions $\Psi_{\pm} \rightarrow 0$ as
$a \rightarrow 0$. The form of the wavefunctions
also guarantees the Hermitivity of the Wheeler-DeWitt
equation \cite{peleg2}
\begin{equation}
J_{+ -} (0) = \frac{i}{2}
\Bigl( \Psi_+^* \frac{\partial}{\partial a} \Psi_-
- \Psi_- \frac{\partial}{\partial a} \Psi_+^*
\Bigr)_{\vert a = 0} = 0.
\end{equation}
{}From Eq. (\ref{mom})
\begin{equation}
m_P \dot{a} = \pm \frac{4 \pi \hbar \sqrt{D_{\nu}}}{3} \frac{1}{a^2},
\end{equation}
we recover exactly the same solutions (\ref{exact}).

\subsection{ k = 1}

We now turn to the cases of $k = \pm 1$.
The exact wavefunctions of Eq. (\ref{gra})
are classified according to the the signs of $D_{\nu}$, as in the $k= 0$ case.
First, for $D_{\nu} > 0$ the wavefunctions are
\begin{equation}
\Psi^{(\pm)}_{III} = C_{\pm} a^{(1 - p)/2}
H^{(1,2)}_{i\sqrt{D_{\nu}}/2} (i \beta a^2)
\end{equation}
whereas for $D_{\nu} < 0$ they are
\begin{equation}
\Psi^{(\pm)}_{IV} = C_{\pm} a^{(1 - p)/2}
H^{(1,2)}_{\sqrt{-D_{\nu}}/2} ( i\beta a^2)
\end{equation}
where $H^{(1,2)}$ are the Hankel functions, and
$\beta = \frac{3m_P}{8 \pi \hbar}$.
For $a >> 1$, the wavefunctions are approximated as
\begin{eqnarray}
\Psi^{(\pm)}_{III} &=& C_{\pm} \sqrt{\frac{2}{\pi \beta a^2}}
e^{\mp  \bigl( \beta a^2 - \frac{\pi \nu}{2} +i \frac{\pi}{4} \bigr)},
\nonumber\\
\Psi^{(\pm)}_{IV} &=& C_{\pm} \sqrt{\frac{2}{\pi \beta a^2}}
e^{\mp  \bigl( \beta a^2 + i \frac{\pi \nu}{2} +i \frac{\pi}{4} \bigr)}.
\end{eqnarray}

Regardless of the signs of $D_{\nu}$, the asymptotic wavefunctions
$\Psi^{(\pm)}_{III}$ and $\Psi^{(\pm)}_{IV}$ describe the
exponentially decreasing and the exponentially
increasing wavefunctions dominated by the curvature, of which
$\Psi^{(+)}_{III}$ and $\Psi^{(+)}_{IV}$ can be interpreted as
the quantum wormholes \cite{page}.
{}From the series expansion of the Hankel functions one can easily see that
the approximate wavefunctions for $ a << 1$ are the same as those
of $k = 0$. We may use these wavefunctions to calculate the
quantum potential and treat the semiclassical Einstein equation.
Thus, we find that there is still a particular class of
solutions, quantum mechanical or semiclassical, which may avoid
classical singularities due to the quantum potential.

\subsection{k = -1}

We find the exact solutions for $k = -1$.
For $D_{\nu} > 0$ the wavefunctions are
\begin{equation}
\Psi^{(\pm)}_{V} = C_{\pm} a^{(1 - p)/2}
H^{(1,2)}_{i\sqrt{D_{\nu}}/2} (\beta a^2)
\end{equation}
whereas for $D_{\nu} < 0$ they are
\begin{equation}
\Psi^{(\pm)}_{VI} = C_{\pm} a^{(1 - p)/2}
H^{(1,2)}_{\sqrt{- D_{\nu}}/2} ( \beta a^2)
\end{equation}
For $a >> 1$, we approximate the wavefunctions as
\begin{eqnarray}
\Psi^{(\pm)}_{V} &=& C_{\pm} \sqrt{\frac{2}{\pi \beta a^2}}
e^{\pm i \bigl( \beta a^2 - i \frac{\pi \nu}{2} - \frac{\pi}{4} \bigr)},
\nonumber\\
\Psi^{(\pm)}_{VI} &=& C_{\pm} \sqrt{\frac{2}{\pi \beta a^2}}
e^{\pm i \bigl( \beta a^2 -  \frac{\pi \nu}{2} - \frac{\pi}{4} \bigr)},
\end{eqnarray}

The asymptotic wavefunctions $\Psi^{(\pm)}_{V}$ and $\Psi^{(\pm)}_{VI}$
for all $D_{\nu}$, represent  either
the expanding or the recollapsing Universe.
For $ a << 1$ the wavefunctions have the same form
as those for $k = 0, 1$. The quantum potential still avoids
the cosmological singularities.

\section{Conclusion}

We applied the de Broglie-Bohm interpretation to the Wheeler-DeWitt equation
for the FRW model with a massless scalar field. We not only found the
exact wavefunctions but also solved both the Hamilton-Jacob equation
with the quantum potential (real part) and the continuity equation for
probability (imaginary part) at the same time. Both the exact wavefunctions
and the de Broglie-Bohm interpretation gave the identical result,
as expected, that there can be or cannot be quantum potential dominated
particular solutions that avoid either the initial or the final cosmological
singularities,
depending on the relative magnitude of the operator ordering
parameter and the momentum of massless scalar field.
It was shown that the quantum potential and the back-reaction of
the massive scalar field changed significantly the effective energy density
in the semiclassical gravity \cite{kim2}.
These results suggest that the quantum potential and the back-reaction
of both geometry and matter fields might change the properties of
and avoid the initial and the final cosmological
singularities.

\vspace{1.5ex}
\begin{flushleft}
{\large\bf Acknowledgements}
\end{flushleft}

This work was supported in parts by Non-Directed Research Fund,
Korea Research Foundation, 1996, and by the Center for Theoretical
Physics, Seoul National University.

\end{document}